\documentclass[12pt,a4paper,twoside]{article}

\usepackage{epsfig}

\newlength{\figheight}
\newlength{\doublefigwidth}
\newlength{\quaterfigwidth}
\def\gg{$\gamma \gamma$}

\def\ggg{$\Gamma_{\gamma \gamma}$}

\def\pgg{$\phi_{\gamma \gamma}$}

\def\epem{$e^+ e^-$}

\def\z0{Z}

\def\mh{M_{\cal H}}
\def\th{\Theta_{\cal H}}

\begin{document}

\title{Model-Independent Determination of CP Violation \\
from Angular Distributions in Higgs Boson Decays \\
to WW and ZZ at the Photon Collider }

\author{
P.Nie\.zurawski, A.F.\.Zarnecki \\
   {\small\it Institute of Experimental Physics, Warsaw University
                Ho\.za 69, 00-681 Warszawa, Poland }
\and  M.Krawczyk \\
   {\small\it Institute of Theoretical Physics, Warsaw University
                Ho\.za 69, 00-681 Warszawa, Poland }
}

\maketitle

\begin{abstract}
The model-independent determination 
of the Higgs-boson CP properties at 
the Photon Collider at TESLA has been studied in detail,
for masses between 200 and 350 GeV,
using realistic luminosity spectra and 
detector simulation.
We consider a generic model with 
the CP violating Higgs tensor couplings 
to gauge bosons. 
We introduce a new variable describing angular distributions 
of the secondary WW and ZZ decay products which is very sensitive 
to the CP properties of the Higgs-boson.
Understanding of the detector performance turns out to be crucial, 
as the influence of the acceptance corrections
is similar to the effects of  CP violation.
From the combined measurement of invariant mass distributions 
and various angular distributions 
the phase describing a CP violation can be determined to about 50~mrad
after one year of Photon Collider running.
\end{abstract}

\section{Introduction}
\label{sec:intro}

The physics potential of a Photon Collider~\cite{tdr_pc} is very
rich and complementary to the physics program of the \epem\
and hadron-hadron colliders.
It is an ideal place to study the mechanism of the electroweak 
symmetry breaking (EWSB) and the properties of the Higgs-boson.

In paper \cite{nzk_wwzz} we performed a realistic simulation of 
the Standard Model (SM) Higgs-boson production at the Photon Collider
for $W^+ W^-$ and $\z0 \z0$ decay channels,
for Higgs-boson masses above 150~GeV.
From the combined analysis of  $W^+ W^-$ and $\z0 \z0$ 
invariant mass distributions
the $\gamma \gamma$ partial width of the Higgs boson, \ggg,
can be measured with an accuracy of 3 to 8\%
and the  phase of $\gamma \gamma \rightarrow h $ amplitude, \pgg, 
with an accuracy between 30 and 100~mrad.
In paper \cite{nzk_smlike} we extended this analysis 
to the generalised Standard Model-like scenario $B_h$
of the Two Higgs Doublet Model II, 2HDM(II), with and 
without CP-violation.
In case of the solution $B_h$ with a weak CP violation
via $H-A$ mixing, the mixing angle $\Phi_{HA}$ can be constrained to 
about 100~mrad for low values of $\tan\beta$. 
We found that for a general 2HDM~(II) with CP violation,
only the combined analysis of LHC, LC and Photon Collider measurements
allows for the precise determination of Higgs-boson 
couplings and of CP-violating $H-A$ mixing angle \cite{nzk_2hdm}.

In this paper our aim is to establish CP-property
of the Higgs-boson in a generic model with a direct CP-violation.
The measurement of the Higgs-boson properties at 
the Photon Collider at TESLA is studied in detail for masses 
between 200 and 350~GeV,
using realistic luminosity spectra and detector simulation. 
The  model with generic, CP-violating Higgs-boson couplings  
to vector bosons \cite{miller} leads to different  
angular distributions for a scalar- and 
pseudoscalar-type of  couplings.  
From  a combined analysis of invariant mass 
distributions and angular distributions 
of the  $W^+ W^-$ and $ZZ$ decay-products
the CP-parity of the observed Higgs state can be determined
independently on a production mechanism \cite{cpang}.
Results given in this paper supersede results 
presented in the second part of our earlier work~\cite{eps2003}.

%
%

\section{Generic Higgs model with CP violating couplings}
\label{sec:model}

Following the analysis described in \cite{miller}
we consider a generic model with a direct CP violation,
with tensor couplings of a Higgs boson, $\cal H$,  to $ZZ$ and $W^+W^-$
given by:
\begin{eqnarray}
g_{{\cal H}ZZ} & = & i g \frac{M_Z}{\cos \theta_W} \; \left(
  \lambda_H \cdot g^{\mu \nu} \; + \; 
  \lambda_A \cdot \varepsilon^{\mu \nu \rho \sigma}\; 
  \frac{(p_1+p_2)_\rho \; (p_1-p_2)_\sigma}{M_Z^2} \right) \; , \nonumber \\
g_{{\cal H}WW} & = & i g M_W \; \left(
  \lambda_H \cdot g^{\mu \nu} \; + \; 
  \lambda_A \cdot \varepsilon^{\mu \nu \rho \sigma}\; 
  \frac{(p_1+p_2)_\rho \; (p_1-p_2)_\sigma}{M_W^2} \right) \; ,
\end{eqnarray}
where $p_1$ and $p_2$ are the 4-momenta of the vector bosons.
Contributions to gauge couplings with $\lambda_H$  have 
a structure of the CP-even SM Higgs boson coupling,\footnote{
Other possible  CP-even tensor structure,  $(p_1+p_2)^{\mu} (p_1+p_2)^{\nu}$,
give the angular distributions similar to that of the SM Higgs boson
and therefore we will not consider this case separately.
}
whereas the ones  with  $\lambda_A$ correspond to
a general CP-odd coupling for the spin-0 boson.
Coefficients $\lambda_H$ and $\lambda_A$ can be written as:
\begin{eqnarray}
\lambda_H & = & \lambda \cdot \cos \Phi_{CP} \; , \nonumber \\*
\lambda_A & = & \lambda \cdot \sin \Phi_{CP} \; .
\end{eqnarray}
The couplings of the Standard Model Higgs boson are reproduced for 
$\lambda = 1$ and $\Phi_{CP} = 0$ (i.e. $\lambda_H = 1$ and $\lambda_A = 0$).
Below we will limit ourselves to  $\lambda \approx 1$ and $|\Phi_{CP}|\ll~1$
region, corresponding to small deviation from the Standard Model
coupling.
However,
we do not make any assumptions concerning Higgs-boson couplings to fermions
and we allow for deviations from SM predictions in \ggg\
and \pgg.
Therefore our results do not depend on the assumed Higgs-boson production
mechanism and our approach can be considered as model-independent.
To simplify the analysis, we only assume that the ${\cal H}$ branching ratios
to $W^+W^-$ and $ZZ$ are the same as in the Standard Model.
No substantial deviations in the $W^+W^-$
and $ZZ$ branching ratios are expected unless Higgs-boson
Yukawa couplings to up- or down-type fermions become very large. 
Possible small deviation in the branching ratios 
are equivalent to \ggg\ variation.


\section{Angular distributions for secondary decay products}

\begin{figure}[p]
  \begin{center}
     \epsfig{figure=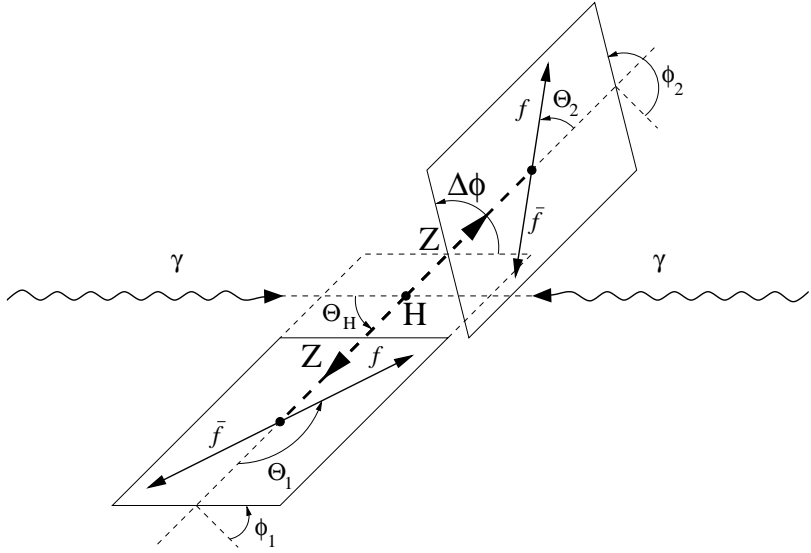,height=\figheight,clip=}
  \end{center}
 \caption{
    The definition of the polar angles $\th$, $\Theta_1$ and $\Theta_2$,
  and the azimuthal angles $\phi_1$ and $\phi_2$ for the process
$\gamma \gamma \rightarrow h \rightarrow Z Z \rightarrow 4\; f $.
$\Delta \phi$ is the angle between two Z decay planes,
$\Delta \phi = \phi_2 - \phi_1$. 
All polar angles are calculated in the rest frame of the decaying particle.
           } 
 \label{fig:ang_def} 
 \end{figure}

Here we consider the Higgs boson decays to gauge bosons.
The angular distributions of the secondary $W^+W^-$
and $ZZ$ decay products turn out to be very sensitive 
to the CP properties of the Higgs-boson \cite{miller}. 
Angular variables which can be used
in the analysis  are defined in Fig.~\ref{fig:ang_def}.
%
%
To test CP-properties of the Higgs-bosons the 
distributions of the polar  angles $\Theta_1$ and $\Theta_2$
as well as the $\Delta \phi$ distribution, where $\Delta \phi$ 
is the angle between two $Z$- or two $W$-decay planes, are used.
Here we propose to consider,
instead of two-dimensional distribution in ($\cos \Theta_1$, $\cos \Theta_2$)
the distribution in a new variable, defined as
\begin{eqnarray}
 \zeta & = & \frac{\sin^2\Theta_1\;\cdot\; \sin^2\Theta_2}
                 {(1+\cos^2\Theta_1)\cdot (1+\cos^2\Theta_2)} \; .
\end{eqnarray}
The $\zeta$-variable  corresponds to the ratio 
of the angular distributions expected for the decay of a scalar 
and a pseudoscalar  (in a limit $\mh >\!\!> M_Z$) \cite{miller}.
It proves to be very useful and complementary to $\Delta \phi$
variable.

\begin{figure}[p]
  \begin{center}
     \epsfig{figure=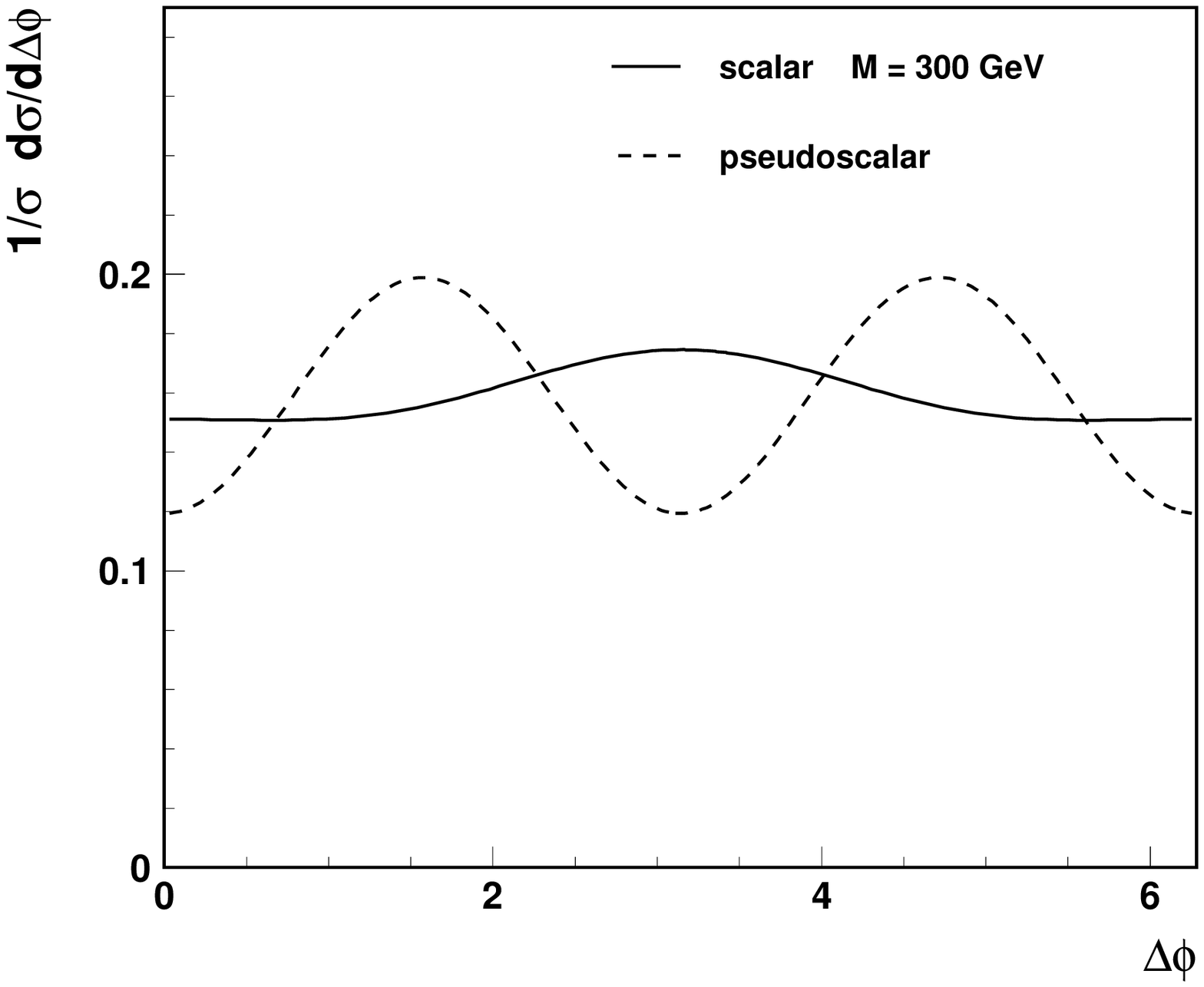,width=\doublefigwidth,clip=}
     \epsfig{figure=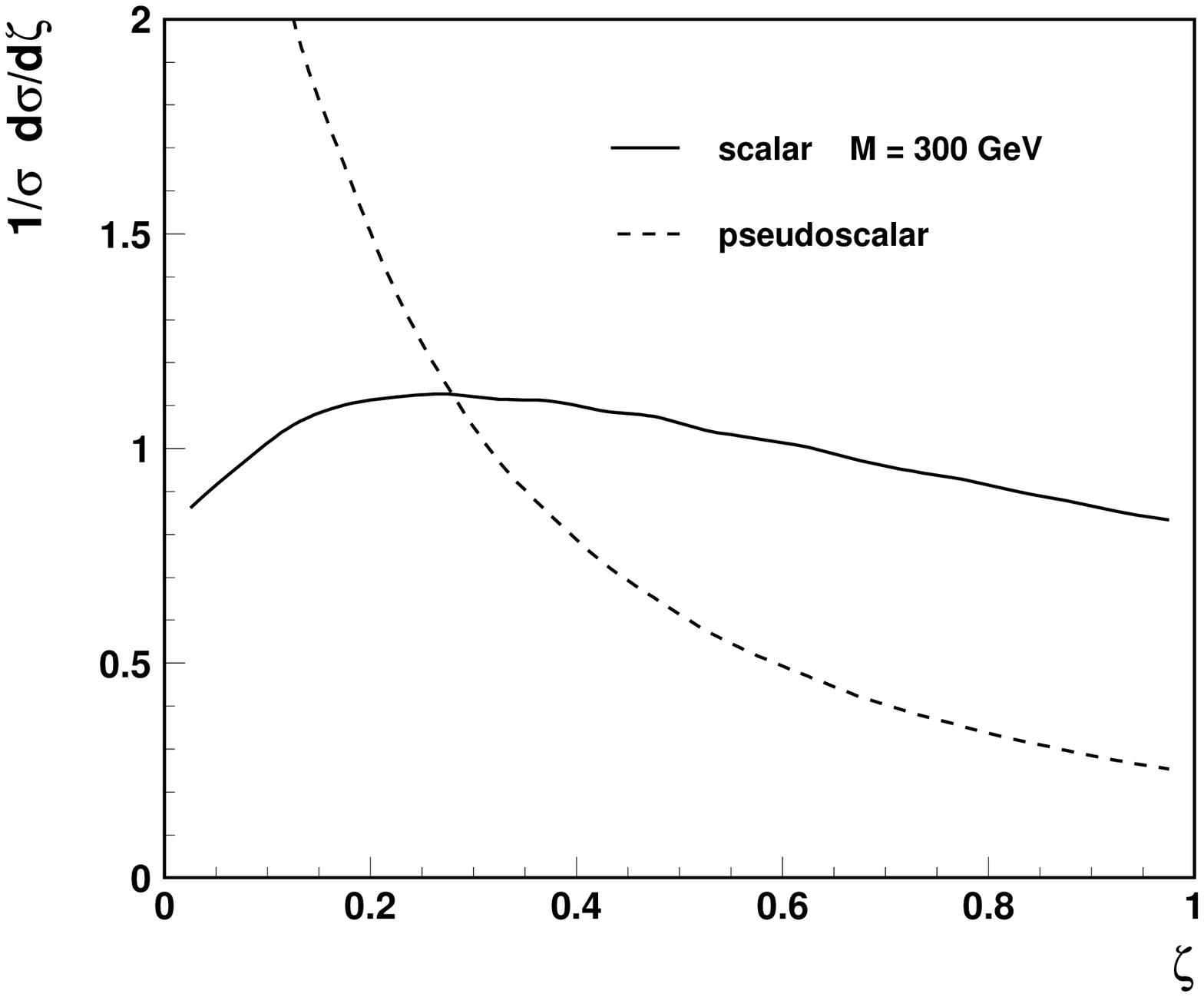,width=\doublefigwidth,clip=}
  \end{center}
 \caption{
Normalised angular distributions in $\Delta \phi$ (left plot) and 
$\zeta$ (right plot), expected 
for scalar and pseudoscalar higgs decays 
$H,A \rightarrow Z Z \rightarrow l^+ l^- j j $,
for the  higgs mass of 300 GeV.
           } 
\label{fig:ang_exp} 
 \end{figure} 

The angular distributions in $\Delta \phi$ and $\zeta$, expected for
a scalar ($\Phi_{CP}=0$) and a pseudoscalar ($\Phi_{CP}=\frac{\pi}{2}$) 
higgs decays, 
\mbox{${\cal H} \rightarrow Z Z \rightarrow l^+ l^- j j $},
are compared in Fig.~\ref{fig:ang_exp}.
They clearly distinguish between decays of scalar ($H$)
and pseudoscalar ($A$) higgs.
For the scalar  higgs the distributions are almost flat in both
$\Delta \phi$ and $\zeta$.
For the pseudoscalar case these two distributions
have totally different character.
From the measurement of one or other, or both of
these distributions it's possible to establish the
CP properties of the Higgs boson, 
even without taking into account the production mechanism.

Note, that the theoretical distribution of the higgs 
decay-angle $\th$ is expected 
to be flat both for scalar and pseudoscalar higgses,
as this is characteristic to spin 0 particle.
However, the measured $\th$ distributions for $H$ and $A$ 
decays can differ due to the different acceptance corrections and
interference with a non-resonant background. 
Therefore, in  the observed $\th$ distributions, 
as well as reconstructed \gg\ invariant mass distributions 
for $W^+W^-$ and $ZZ$ decays, some sensitivity to the 
Higgs-boson couplings can show up.


\section{Reconstruction of the angular distributions}

Our analysis uses the CompAZ parametrisation \cite{compaz} of
the realistic luminosity spectra for a Photon Collider 
at TESLA.
We assume that the centre-of-mass energy of colliding electron beams, 
$\sqrt{s_{ee}}$, is optimised for the production of
a Higgs boson with a given mass.
All results presented in this paper were obtained 
for an integrated luminosity corresponding to one year 
of the photon collider running.
The total photon-photon luminosity increases from about
600~fb$^{-1}$ for $\sqrt{s_{ee}}=305$~GeV 
(optimal beam energy choice for $\mh=200$~GeV) to
about 1000~fb$^{-1}$ for $\sqrt{s_{ee}}=500$~GeV
($\mh=350$~GeV).

\begin{figure}[p]
\begin{center}
\epsfig{figure=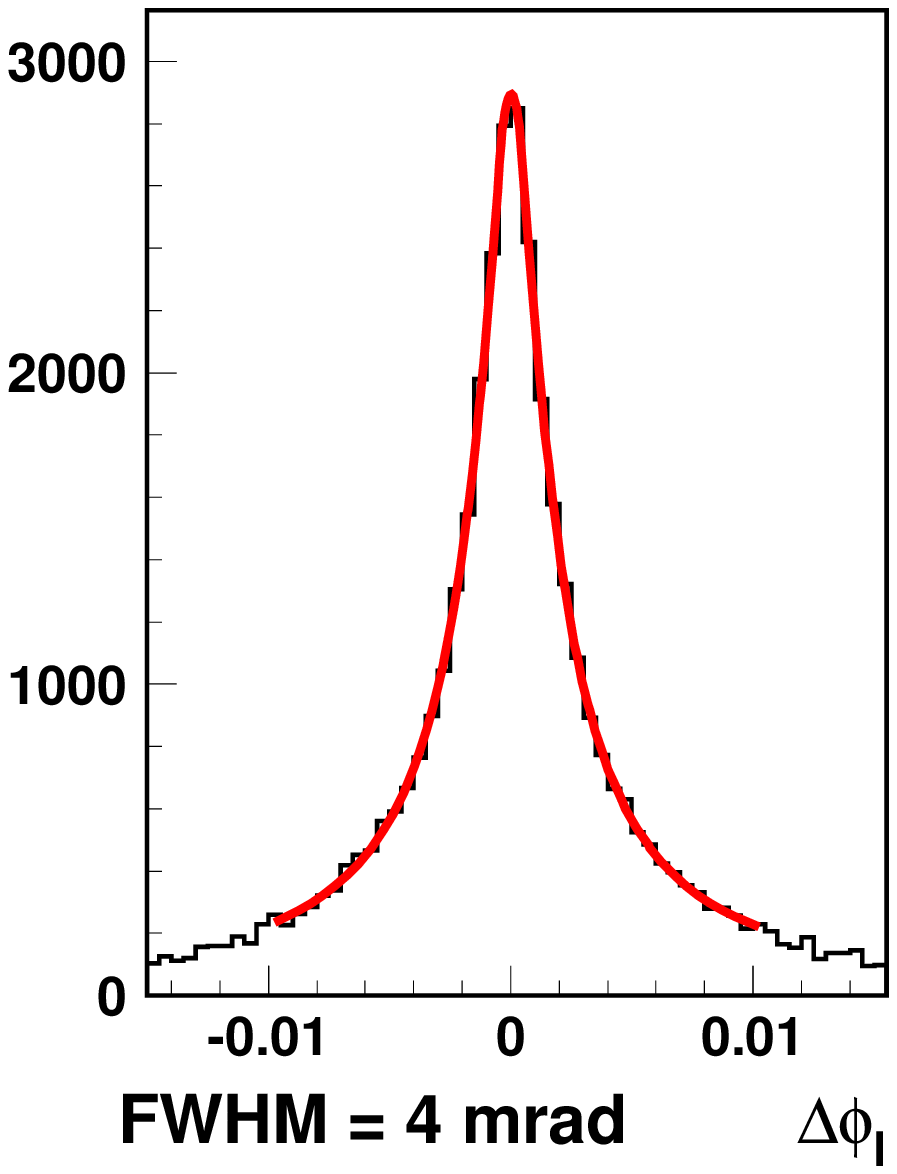,width=\quaterfigwidth,clip=}
\epsfig{figure=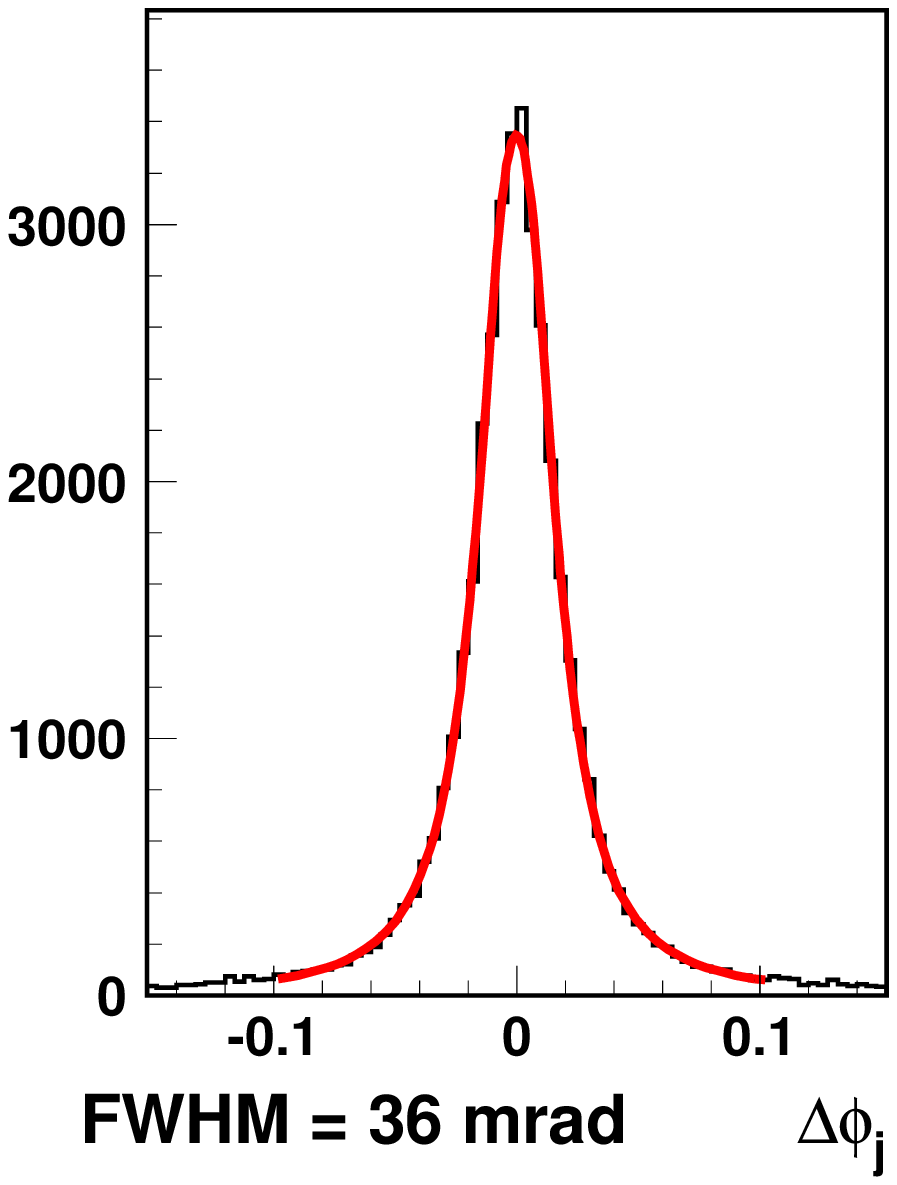,width=\quaterfigwidth,clip=}
\epsfig{figure=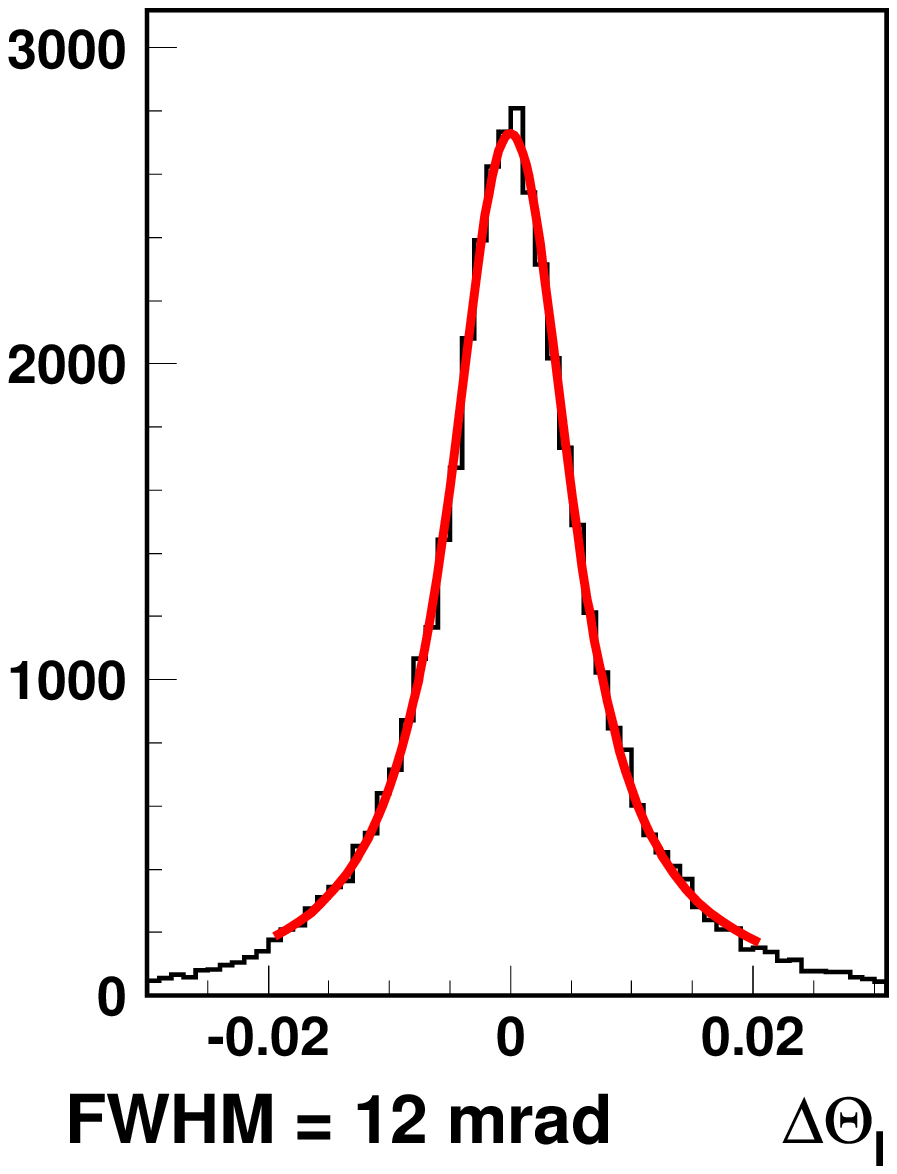,width=\quaterfigwidth,clip=}
\epsfig{figure=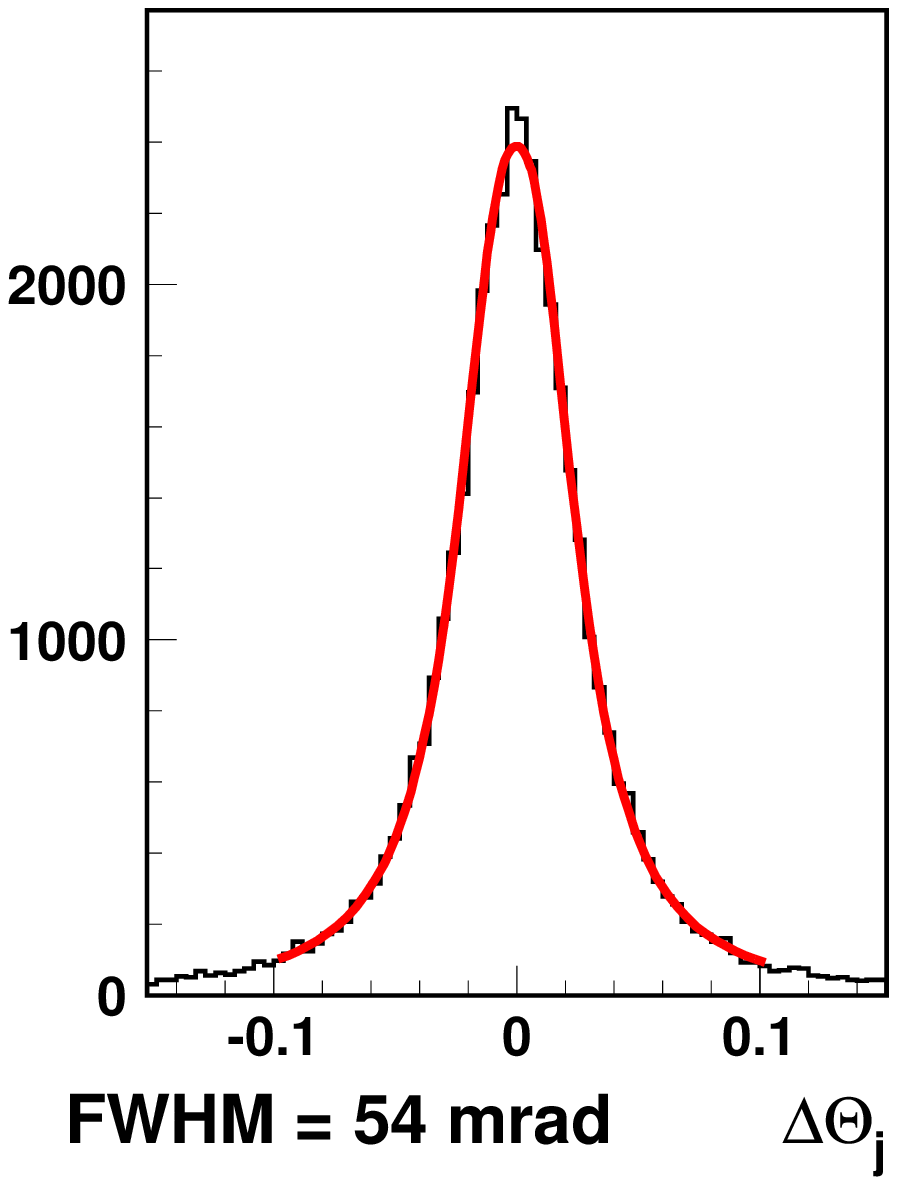,width=\quaterfigwidth,clip=}
\end{center}
 \caption{
     Resolution in the reconstructed $Z$-decay angles $\Theta$ and $\phi$, 
  for the leptonic ($\Theta_l$, $\phi_l$)  and hadronic ($\Theta_j$, $\phi_j$)
     final states.
           Events were simulated with the PYTHIA and SIMDET programs, for
           a primary electron-beam energy of 209 GeV and 
          Standard Model Higgs-boson mass of 300~GeV.
           } 
 \label{fig:ang_res} 
 \end{figure}

Measurement of the angular distributions has been studied 
using the samples of $ZZ$ events generated with the PYTHIA~\cite{PYTHIA} 
and SIMDET~\cite{SIMDET} programs, as described in \cite{nzk_wwzz}.
The resolutions of the reconstructed decay angles $\Theta$ and $\phi$, 
for the scalar Higgs-boson with mass of 300~GeV
(primary electron-beam energy of 209~GeV)
are compared in Fig.~\ref{fig:ang_res}
for the leptonic  and hadronic $Z$ decays.

\begin{figure}[p]
  \begin{center}
     \epsfig{figure=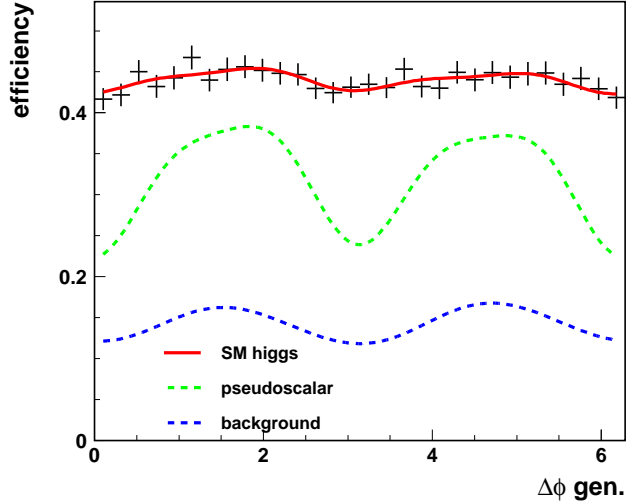,height=\figheight,clip=}
  \end{center}
 \caption{ 
         Selection efficiency as a function of the angle $\Delta \phi$
        between two $Z$ decay planes, for 
        $Z Z \rightarrow l^+ l^- j j $ events
         coming from the scalar higgs decays, pseudoscalar higgs decays 
         and non-resonant background.
         Events were simulated with the PYTHIA and SIMDET programs, for
         a primary electron-beam energy of 209 GeV and 
         Higgs-boson mass of 300~GeV.
      Only events with the reconstructed higgs mass between 286 and  312 GeV
         are accepted.
         } 
 \label{fig:acc_300_705} 
 \end{figure}

A very good resolution is obtained for all considered distributions,
nevertheless the measured angular distributions are strongly affected by 
the selection cuts used in the analysis.
Because of the cut on the lepton and jet angles, applied
to preserve a good mass resolution (see \cite{nzk_wwzz} for more details),
we observe a significant loss of the selection efficiency for
the events with lepton or jet emitted along the beam direction.
Also the Durham jet algorithm used in the event reconstruction
imposes constraints on the angular separation between leptons and jets.
Both effects result in a highly non-uniform angular acceptance.
The obtained selection efficiencies 
for $\gamma \gamma \rightarrow Z Z \rightarrow l^+ l^- j j $ events,
as a function of the angle $\Delta \phi$,
are presented in Fig.~\ref{fig:acc_300_705}.
Additional cuts on the reconstructed $ZZ$ invariant-mass have been
introduced to optimise the signal measurement:
for Higgs boson mass of 300~GeV the accepted mass range lies 
between 286 and  312~GeV.
The efficiencies for $ZZ$ events coming from  decays of scalar, 
pseudoscalar higgs and non-resonant background are compared.
The polar-angle distributions differ among these three classes of events,
therefore significantly different acceptances are obtained.
Understanding of this effect is crucial,
as it can mimic a pseudoscalar type of distribution.
For the first time such effect is taken into account 
in the analysis of the considered process.

In the following analysis, the reconstructed $\Delta \phi$ values 
range from 0 to $\pi$, since we are not able to distinguish 
between quark and antiquark jet.
%
%
%

\section{Analysis}

For arbitrary values of model parameters $\lambda$ and $\Phi_{CP}$
we calculate the expected angular and invariant mass distributions 
for $ZZ$ and $W^+ W^-$ events by convoluting the corresponding 
cross-section formula with the analytic photon-energy spectra
CompAZ~\cite{compaz}. 
To take into account detector effects, 
we convolute this further with the function parameterising 
the invariant-mass resolution and the acceptance 
function containing  the angular- and jet-selection cuts.%
\footnote{To simplify acceptance calculations, 
we neglect effects of the finite angular resolution.}
For the measurement of the event distributions in 
$\Delta \phi$ and $\zeta$ we introduce the  additional cuts 
on the reconstructed $ZZ$ invariant mass (for $ZZ$ events)
or on the reconstructed  $W^+ W^-$ invariant mass as well as on 
the higgs-decay angle $\th$ (for $W^+ W^-$ events).
The cuts were optimised for the smallest relative error in 
the signal cross-section measurement.

%
\begin{figure}[p]
  \begin{center}
     \epsfig{figure=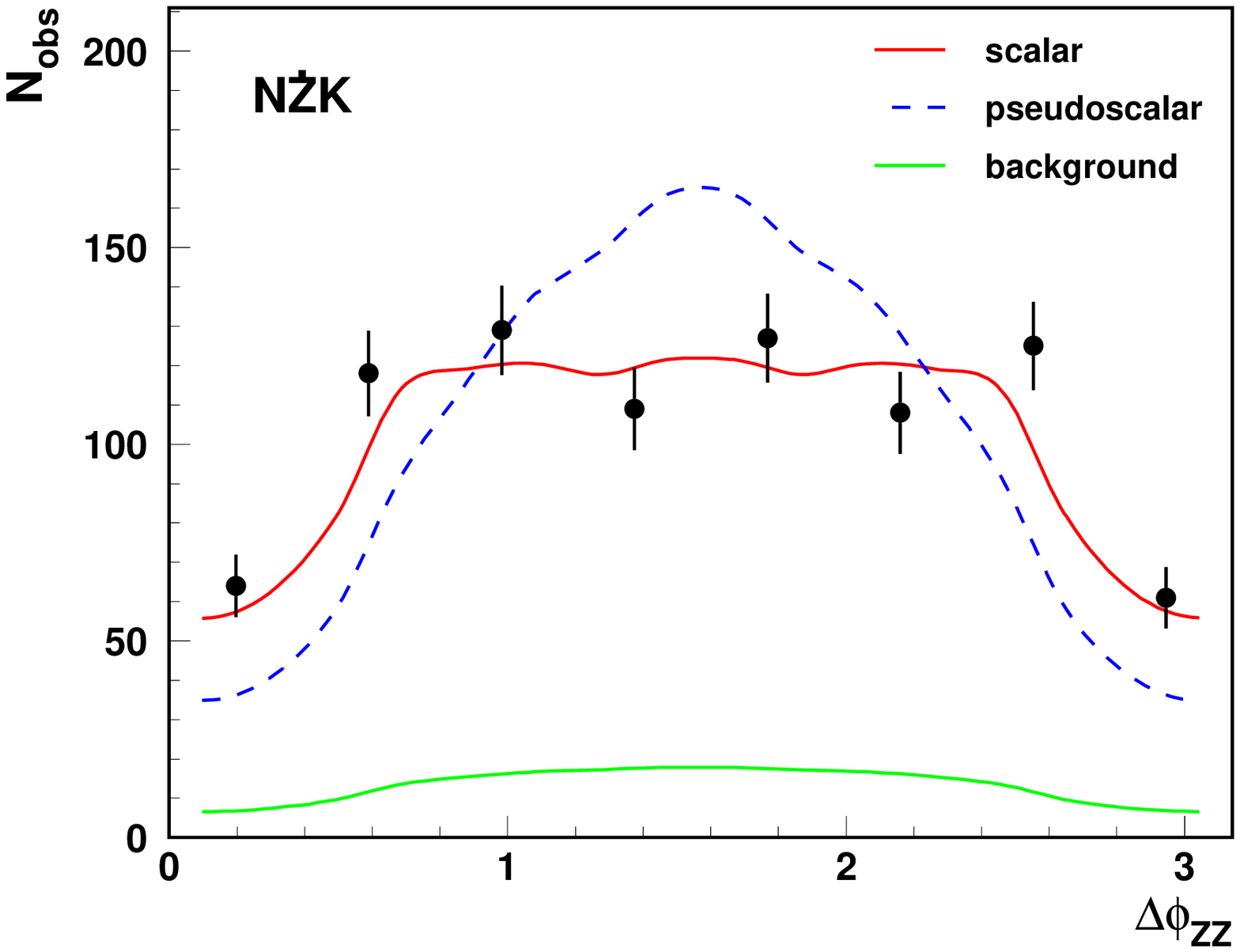,width=\doublefigwidth,clip=}
     \epsfig{figure=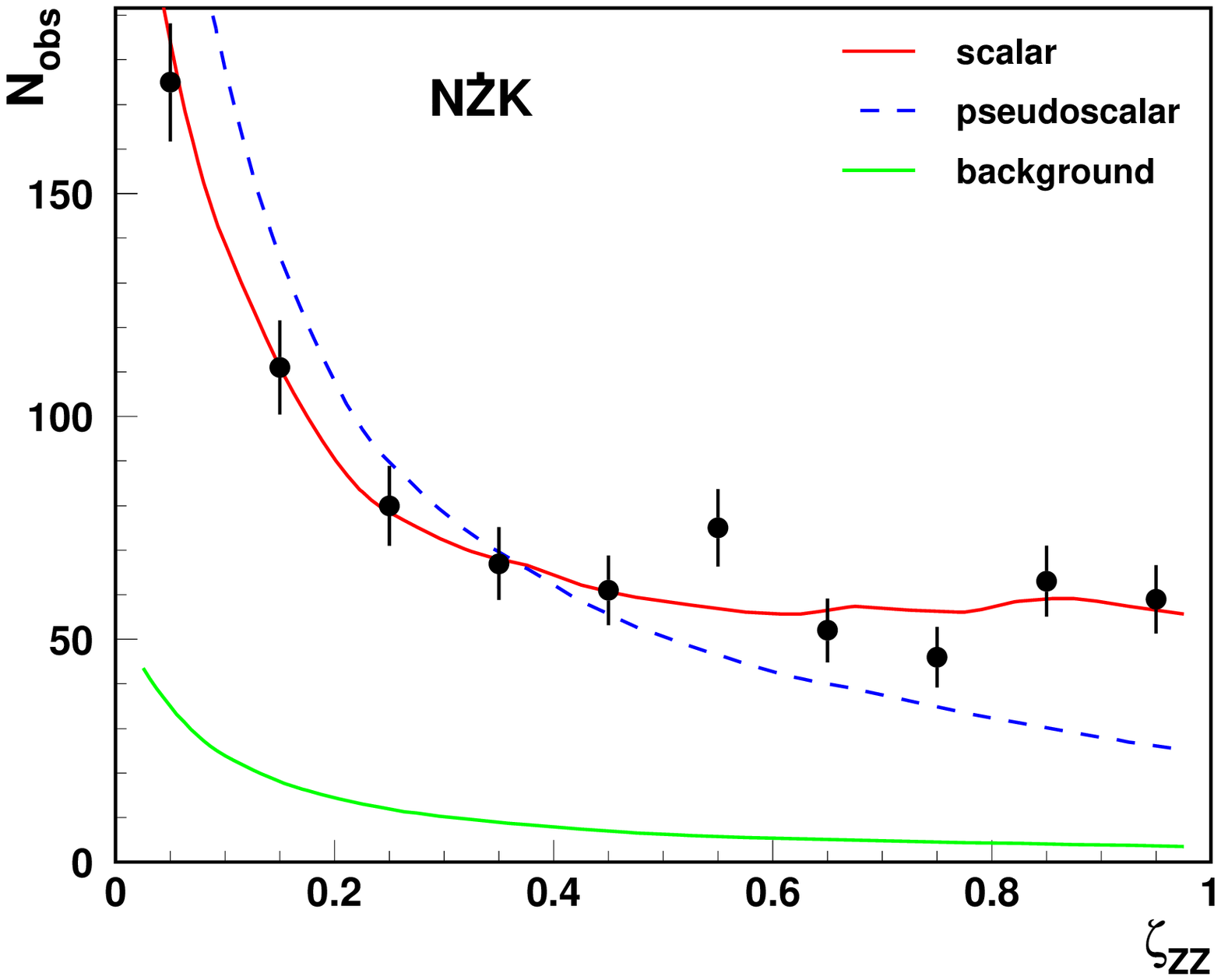,width=\doublefigwidth,clip=}
  \end{center}
 \caption{ 
        Measurement of the angle $\Delta \phi_{ZZ}$
        between two $Z$-decay planes (left plot) and
        of the variable $\zeta_{ZZ}$ calculated
        from the polar angles of the $Z\rightarrow l^+ l^-$
        and  $Z\rightarrow j j $  decays (right plot)
        for  $Z Z \rightarrow l^+ l^- j j $ events.
         Points with error bars indicate the statistical precision of the
         measurement after one year of Photon Collider
        running at nominal luminosity. 
         The solid (red) and dashed (blue) lines correspond to predictions
         of the model with pure scalar ($\Phi_{CP}=0$) and
         pseudoscalar ($\Phi_{CP}=\frac{\pi}{2}$) 
         Higgs-boson couplings, respectively.
         Green line represents the SM background of non-resonant
         $ZZ$ production.         
         Signal and background calculations are performed for
         primary electron-beam energy of 152.5~GeV and 
         the Higgs-boson mass of 200~GeV.
         } 
 \label{fig:histzz} 
 \end{figure} 
%
%
\begin{figure}[p]
  \begin{center}
     \epsfig{figure=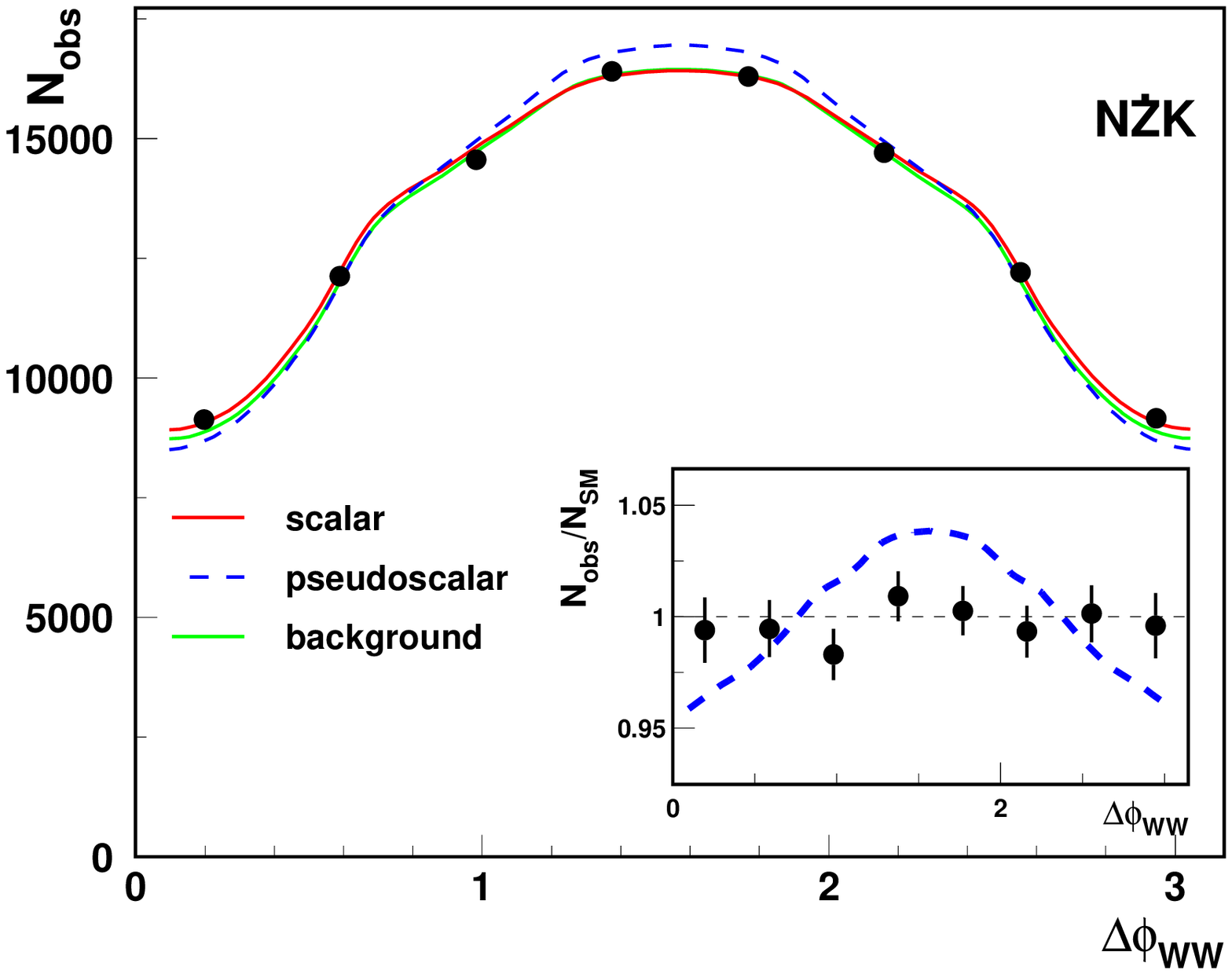,width=\doublefigwidth,clip=}
     \epsfig{figure=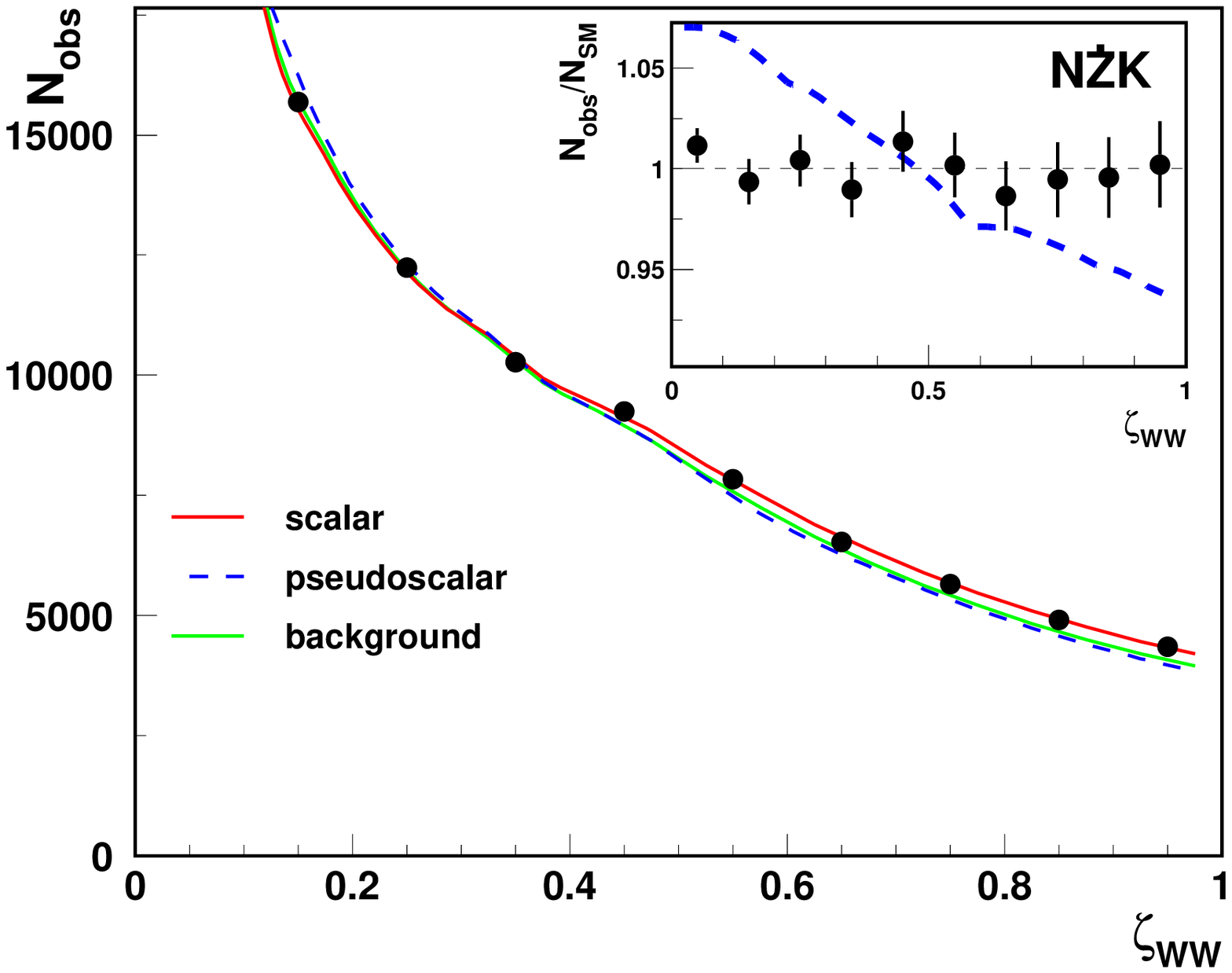,width=\doublefigwidth,clip=}
  \end{center}
 \caption{ 
        Measurement of the angle $\Delta \phi_{WW}$
        between two $W$-decay planes (left plot) and
        of the variable $\zeta_{WW}$ calculated
        from the polar angles of two $W\rightarrow j j$
        decays (right plot)
        for  $WW \rightarrow 4 j $ events.
        Points with error bars indicate the statistical precision of the
        measurement after one year of Photon Collider
        running at nominal luminosity. 
         The solid (red)  and dashed (blue) lines correspond to predictions
         of the model with pure scalar ($\Phi_{CP}=0$) and
         pseudoscalar ($\Phi_{CP}=\frac{\pi}{2}$) 
         Higgs-boson couplings, respectively.
         Green line represents the SM background of non-resonant
         $ZZ$ production.         
         Signal and background calculations are performed for
         primary electron-beam energy of 152.5~GeV and 
         the Higgs-boson mass of 200~GeV.
         The insets show the relative deviations from 
         the Standard Model predictions ($\Phi_{CP}=0$)
         for pseudoscalar  Higgs-boson couplings.
         } 
 \label{fig:histww} 
 \end{figure} 
%
%

The expected precision in the measurements of the $\Delta \phi$-  
and of the  $\zeta$-dis\-tri\-bu\-tions, for 
$\gamma \gamma \rightarrow Z Z \rightarrow l^+ l^- j j $ events
and $\gamma \gamma \rightarrow W^+ W^- \rightarrow 4  j $ events
is illustrated in Figs.~\ref{fig:histzz}  and \ref{fig:histww}, 
respectively. 
Calculations were performed for the
primary electron-beam energy of 152.5~GeV and 
the Higgs-boson mass of 200~GeV.
The results  are compared with the expectation of the generic model with
$\Phi_{CP} = 0$ (as in SM) and $\Phi_{CP} = \frac{\pi}{2}$.
For better comparison of the shape of distributions, 
results for pseudoscalar Higgs-boson
couplings ($\Phi_{CP} = \frac{\pi}{2}$) are normalised to the
Standard Model expectations. 
We see, that even after taking into account beam spectra, 
detector effects, selection cuts and background influence, 
the differences between shapes of the angular distributions
for the scalar and pseudoscalar couplings
are still significant.
Therefore we should be able to constrain Higgs-boson couplings from
the shape of the distributions, even if the overall normalisation
related to the Higgs-boson production mechanism is not known.
For the Standard-Model Higgs-boson decays to $ZZ$, 
about 675 Higgs-boson events and 145 non-resonant
background events are expected after all selection cuts, 
in the selected mass window from 180 to 210~GeV.
For the $WW$ channel about 8000 signal events are expected,
compared to about 170~000 background events.
In both cases a statistical precision of the signal measurement
is similar. 
However, due to overwhelming background contribution,
the $WW$ analysis is much more dependent on systematic uncertainties.


\begin{figure}[b]
  \begin{center}
     \epsfig{figure=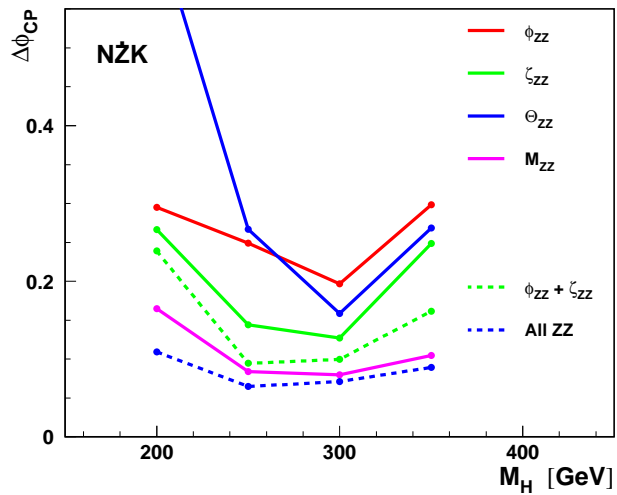,height=\figheight,clip=}
  \end{center}
 \caption{ 
          Statistical error in the determination of 
$\Phi_{CP}$,  expected after one year of Photon Collider running,  
as a function  of the Higgs-boson mass $\mh$.
Fits were performed to the shape of the three individual 
angular distributions measured for $ZZ$ events, as indicated in the plot,
and to the reconstructed invariant mass distribution.
Results of the simultaneous fits to three angular distributions 
and to all distributions are also shown.
All other parameters, except for the overall normalisation,
are  fixed to the Standard Model values.
Errors were calculated assuming $\Phi_{CP} \approx 0$.
         } 
 \label{fig:zzcomp} 
 \end{figure} 


Each of the considered angular distributions discussed above
can be fitted with model expectations, given in terms of 
the parameters $\lambda$ and $\Phi_{CP}$ describing Higgs-boson
couplings to gauge bosons, parameters \ggg\ and \pgg\ describing 
the production mechanism, and an overall normalisation.
Statistical errors in the determination of $\Phi_{CP}$,
resulting from fits to  the different distributions for $ZZ$ events,
are compared in Fig.~\ref{fig:zzcomp}.
The remaining model parameters, except for the normalisation, 
are fixed to the Standard Model values.
Out of the three considered angular distributions, 
parameter $\zeta$ turns out to be the most sensitive to the 
angle $\Phi_{CP}$ describing the CP-violation in the Higgs-boson couplings.
Surprisingly, the smallest error is obtained from the fit
to the invariant mass distribution.
This is because,
due to the different selection efficiencies for the scalar and pseudoscalar 
decays and also to interference effects, 
the relative normalisation of the Higgs-boson signal to the non-resonant 
background depends on the angle  $\Phi_{CP}$.
Therefore, we include the invariant mass distributions 
for $W^+W^-$ and $ZZ$ decays in the combined fit.
When all parameters are allowed to vary in such a fit,  
the invariant-mass distributions constrain mainly the \ggg\  and \pgg.

\section{Results}

\begin{figure}[b]
  \begin{center}
     \epsfig{figure=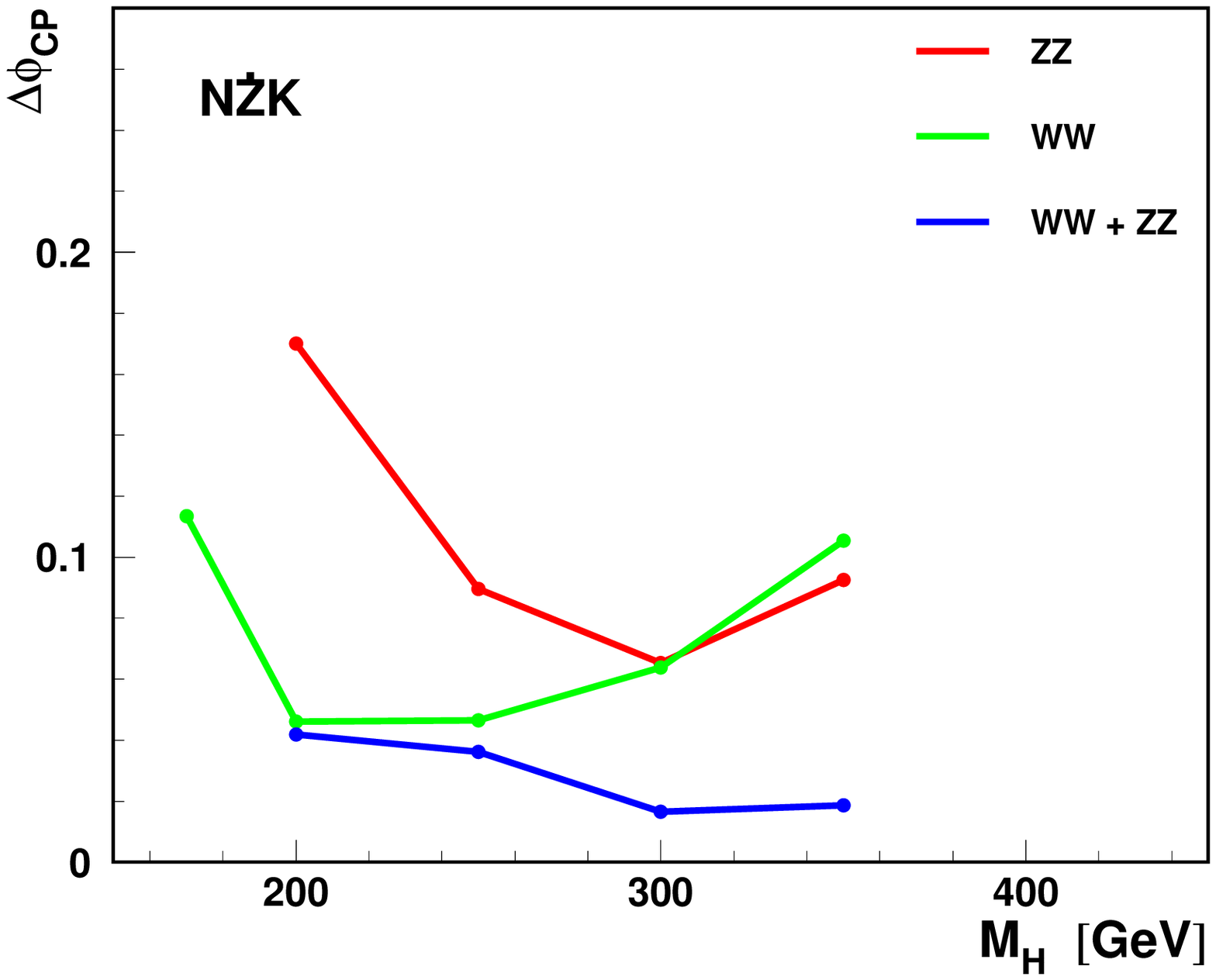,width=\doublefigwidth,clip=}
     \epsfig{figure=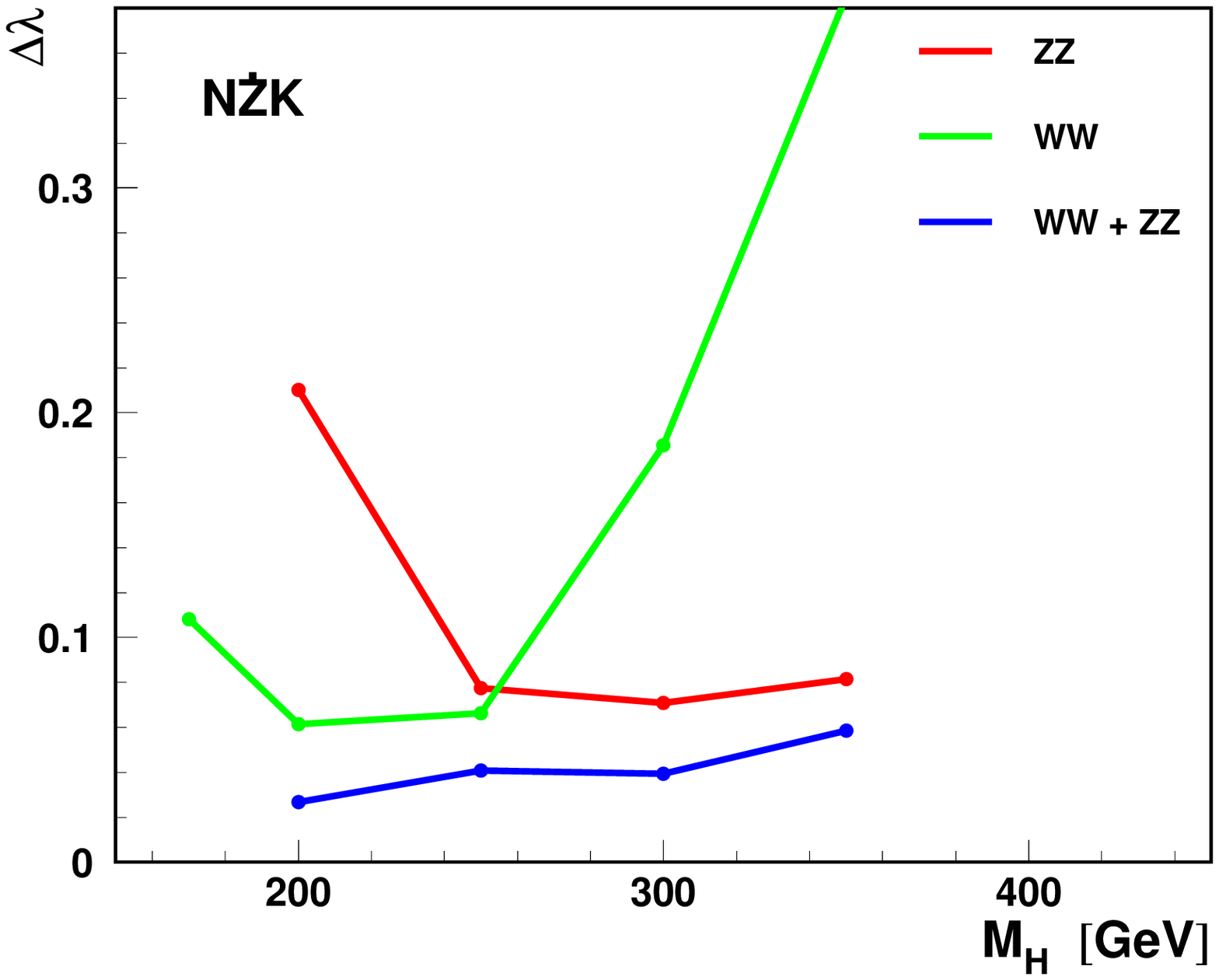,width=\doublefigwidth,clip=}
  \end{center}
 \caption{ 
Statistical error in the determination of $\Phi_{CP}$ (left plot)
and $\lambda$ (right plot),  
expected after one year of Photon Collider running,  
as a function  of the Higgs-boson mass $\mh$.
Combined fits were performed to the considered angular distributions
and invariant mass distributions 
for $ZZ$ events and $W^+ W^-$  events.
Results were obtained assuming small deviations from 
Standard Model predictions, i.e. $\lambda \approx 1$ and
$\Phi_{CP} \approx 0$.
Two photon width of the Higgs boson, \ggg, the phase \pgg\
and normalisations of both samples are allowed to vary in the fit.
         } 
 \label{fig:results} 
 \end{figure}

We calculate the expected statistical errors 
on the parameters $\lambda$ and $\Phi_{CP}$,
from the combined fit to angular distributions measured
for $ZZ$ and $W^+ W^-$  decays, and to the invariant mass distributions.
Results are shown in  Fig.~\ref{fig:results}.
The two photon width of the Higgs boson, \ggg, the phase \pgg\
and normalisations of both samples are allowed to vary in the fit,
so the results are independent to the production mechanism.
One observes that
for low Higgs-boson masses below 250~GeV, 
better constrains are obtained from the measurement of $W^+ W^-$ events, 
whereas for masses above 300~GeV smaller errors are obtained from 
the measurement of $ZZ$  events.
The error on $\Phi_{CP}$ expected from the combined fit 
is below  50~mrad in the whole considered mass range.
The corresponding error on $\lambda$ is about 0.05.

\section{Summary}

An opportunity of measuring the Higgs-boson properties at 
the Photon Collider at TESLA has been studied in detail for masses 
between 200 and 350~GeV,
using realistic luminosity spectra and detector simulation. 
We considered measurement of the invariant mass distributions
and the various angular distributions.
A new variable is proposed to describe the polar angle distributions
of the secondary $W^+ W^-$ and $Z Z$-decay products.
Event reconstruction and selection procedure result
in acceptance corrections which are highly non-uniform and depend
on the Higgs boson CP properties.
Understanding of these effects is crucial in the analysis of angular 
distributions.
From the model-independent study
the angle describing a CP violation in the generic Higgs-boson couplings 
to vector bosons can be determined with accuracy of about 50~mrad.

\subsection*{Acknowledgements}

We would like to thank our colleagues from the 
ECFA/DESY study groups for many useful comments and suggestions.
This work was partially supported 
by the Polish Committee for Scientific Research, 
grant  no.~1~P03B~040~26
and
project no.~115/E-343/SPB/DESY/P-03/DWM517/ 2003-2005.
P.N.~acknowledges a partial
support by Polish Committee for Scientific Research, grant 
no.~2~P03B~128~25.
M.K.~acknowledges a partial
support by the European Community's
Human Potential Programme under contract HPRN-CT-2000-00149 Physics
at Colliders.


\end{document}